\documentclass[aps,prb,twocolumn,showpacs,amssymb,amsmath,superscriptaddress,floatfix]{revtex4}
\usepackage{graphicx}

\newcommand{\ket}[1]{\left|#1\right\rangle}
\newcommand{\bra}[1]{\left\langle #1 \right|}

\begin{document}

\title{Nonequilibrium excitations in Ferromagnetic Nanoparticles}

\author{Silvia Kleff}
\email[]{kleff@theorie.physik.uni-muenchen.de}
\affiliation{Ludwig-Maximilians-Universit\"at, Theresienstr. 37,
               80333 M\"unchen, Germany}
\affiliation{Institut f\"ur Theoretische Festk\"orperphysik,
               Universit\"at Karlsruhe, 76128 Karlsruhe, Germany}
\author{Jan von Delft}
\affiliation{Ludwig-Maximilians-Universit\"at, Theresienstr. 37,
               80333 M\"unchen, Germany}

\date{\today}

\begin{abstract}

  In recent measurements of tunneling transport through individual
  ferromagnetic Co nanograins, Deshmukh, Gu\'eron, Ralph {\em et
    al.}~\cite{mandar,gueron} (DGR) observed a tunneling spectrum with
  discrete resonances, whose spacing was much smaller than what one
  would expect from naive independent-electron estimates. In a
  previous publication, \cite{prl_kleff} we had suggested that this
  was a consequence of nonequilibrium excitations, and had proposed a
  ``minimal model'' for ferromagnetism in nanograins with a discrete
  excitation spectrum as a framework for analyzing the experimental
  data. In the present paper, we provide a detailed analysis of the
  properties of this model: We delineate which many-body electron
  states must be considered when constructing the tunneling spectrum,
  discuss various nonequilibrium scenarios and compare their results
  with the experimental data of Refs.~\onlinecite{mandar,gueron}. We
  show that a combination of nonequilibrium spin- and single-particle
  excitations can account for most of the observed features, in
  particular the abundance of resonances, the resonance spacing and
  the absence of Zeeman splitting.

\end{abstract}

\maketitle

\section{\label{sec:introduction}Introduction}

An important milestone in the study of itinerant ferromagnetism
was reached during the last two years, when
Deshmukh, Gu\'eron, Ralph {\em et al.}~\cite{mandar,gueron} (DGR),
using single-electron tunneling spectroscopy,\cite{RBT}
succeeded for the first time to resolve discrete
resonances in the tunneling spectrum through individual
ferromagnetic single-domain Cobalt nanograins, with diameters
between 1 and 4~nm.
 Their work goes beyond
previous studies of ferromagnetic single electron
transistors,\cite{ono,fert1,fert2,kramer} which elucidated the interplay of
 ferromagnetism and charging effects: the fact that DGR's Co
grains were sufficiently small that discrete resonances could be
resolved, means that they were probing the true quantum states
participating in electron tunneling, which allows the nature of
electron correlations in itinerant ferromagnets to be
studied in unprecedented detail. Besides the intrinsic scientific
interest of studying ferromagnetism on the nm-scale, the insights
so gained might also be of technological interest, since the size
of memory elements in magnetic storage technologies is decreasing
extremely rapidly,~\cite{johnson} and particles as small as 4 nm
are coming under investigation.~\cite{murray}

An examination of the magnetic-field dependence of the individual
resonances observed by DGR indicated that two-current models, in which
spin-up and spin-down electron bands are considered effectively
independently, are inadequate for describing the true electronic
states inside a nm-scale ferromagnet -- so that a fundamentally
different theoretical approach is required.  To this end, a simple new
phenomenological model was recently introduced by the present authors
together with Deshmukh and Ralph,\cite{prl_kleff} and independently by
Canali and MacDonald.\cite{canali} We regard this as a ``minimal
model'' for ferromagnetic nanograins, in that it seems to be the
simplest model possible for a discrete-state system which takes into
account the electronic correlations induced by magnetic interactions
and which treats the particle's total spin as a quantum mechanical
variable.  We argued\cite{prl_kleff} that the main features of the
measured tunneling spectra can be understood within this model by
assuming that  nonequilibrium spin accumulation occurs,
which can produce a much denser spectrum of
tunneling excitations than expected within an independent-electron model.
The conclusion that nonequilibrium effects play an important role has
since been confirmed by more recent measurements by DGR~\cite{mandar}
on a gated device, in which new resonances appeared as the gate voltage
was tuned to drive the system further away from equilibrium.

In the present paper, we provide a detailed analysis of
equilibrium and nonequilibrium tunneling through ferromagnetic grains,
within the framework of our minimal model. The present
analysis goes well beyond that of
Ref.~\onlinecite{prl_kleff}, in that we consider
not only the spin ground states of Fig.~2 in
Ref.~\onlinecite{prl_kleff},
but also spin wave excitations and single-particle excitations.
The latter turn out to be necessary to understand why
a small resonance spacing is observed even when
the threshold bias voltage for the onset of tunneling
is rather small, so that nonequilibrium effects
cannot be very strong.

The structure of the paper is as follows. In
Sec.~\ref{sec:experimental_results} we summarize the experimental
results of Deshmukh {\em et al.}~\cite{mandar} and Gu\'eron {\em
et al.}~\cite{gueron} The Hamiltonian of our minimal model is
presented in Sec.~\ref{sec:model}, together with a
convenient set of basis states for analyzing the low-lying
excitations and their energies. In Sec.~\ref{sec:equilibrium} we
discuss two different equilibrium excitations in ferromagnetic
grains, namely single particle excitations  and spin  excitations.
A detailed discussion of nonequilibrium excitations and their
consequences for tunneling spectra is given in
Sec.~\ref{sec:nonequilibrium}: First we calculate the current
through a grain for a nonequilibrium scenario involving
only transitions between the ground states $|s,s\rangle$ of a
ladder of spin multiplets of different total spin $s$; the
resonance spacing for the peaks in the conductance is found to
be in quantitative agreement with DGR's measurements if we assume
a total ground state spin of about $s_0 \simeq 1000$ and that the
resonances are predominantly due to the tunneling of minority
electrons. We then generalize this nonequilibrium scenario
by including also  all more highly-excited states
$|s,m\rangle$
 of these spin  multiplets, finding that if the total spin is
large $(s \gg 1)$, the Zeeman splitting of the observed
resonances should be strongly suppressed, in agreement with
DGR's experiments. Finally, we show that a combination of
single-particle-
 and spin  excitations in the presence of nonequilibrium can
lead to the large number of resonances seen in experiments. Some
concluding remarks can be found in Sec.~\ref{sec:conclusions}.

A  brief account of our results on the most simple  nonequilibrium
scenario in ferromagnetic grains
has already been given in Ref.~\onlinecite{prl_kleff}.
The present paper includes
a detailed derivation of these results,
(Sec.~\ref{sec:nonequilibrium_spin_ground_states}),
 since  this paves the way for the more complicated
nonequilibrium scenarios presented in
Sec.~\ref{sec:nonequilibrium_spin_wave} and
Sec.~\ref{sec:nonequilibrium_single_particle}, which have not
been reported before.

\section{\label{sec:experimental_results}Summary of experimental results}

In DGR's experiments,~\cite{mandar,gueron} a nanoscale Cobalt grain
 was used as a central island in a single electron transistor:
 it was connected via tunnel barriers to external leads and
 for one of the grains in Ref.~\onlinecite{mandar}, the central grain
  was also capacitively
coupled to a gate. The electronic spectrum of the particle was
determined by measuring the tunnel conductance through the grain
as function of transport
 voltage ($V$), [gate voltage ($V_{\textrm{g}}$)] and magnetic field $\mu_0 H$ at a fixed temperature of
$\lesssim 90$mK.
The diameters of the Co-grains were estimated as 1-4 nm.
Assuming a roughly hemispherical shape,
the number of atoms in each grain then is in
the range $N_{\textrm a} \approx 20$--1500,
implying a total spin of  $s_0 \approx 0.83 N_{\textrm a}
\approx 17$--1250.

Since the charging energy ($>30$meV)  was very much larger
than typical values of the transport voltage ($eV<9$meV)
and the temperature, fluctuations in electron
number on the grain are strongly suppressed, so that
coherent superposition between different electron numbers $N$ need not be considered.
The energy balance condition that determines through which eigenstates
of the grain electrons can tunnel for given values of transport (and gate)
voltage thus involve differences between eigenenergies of a grain with
{\em fixed particle number } $N$ or $N\pm1$,~\cite{jvdralph}

\begin{equation}
\label{energydiff}
\Delta E^{\pm}_{fi} \equiv E_f^{N\pm 1} - E_i^N \, ,
\end{equation}
each corresponding to the energy cost of some rate-limiting electron tunneling
process $|i\rangle^{N} \to |f\rangle^{N \pm 1}$ onto or off the grain.  Here
$|i\rangle^{N}$ denotes a discrete eigenstate, with eigenenergy $E^{N}_i$, of
a grain with $N$ electrons, etc.
As the magnetic field $H$ is swept, the resonances undergo energy shifts and crossings.

    The excitation spectra measured by DGR had several
    properties that differ strikingly from those of previously-studied
    non-magnetic Al and Au grains
    \cite{RBT,drago} including:\\
    (P1): Many \emph{more low-energy excitations} were observed
    than expected: For all Co grains studied, the observed level
    spacing is $d_{\textrm{obs}} \lesssim 0.2$ meV, which is much
    \emph{smaller} than the  independent-electron estimate of
    $d_{\textrm{min}}\approx 1.2\,\textrm{eV}/s_0$ and
    $d_{\textrm{maj}}\approx 4.6\,\textrm{eV}/s_0$ for
    minority-/majority electrons,~\cite{Papa,bulkN} respectively.
    (In Ref.~\onlinecite{mandar} many more resonances were
    observed than in Ref.~\onlinecite{gueron}, due to lower-noise
    tunnel barriers.)
    \\
    (P2): In the small-field regime ($\mu_{0}H<0.2$ T), the tunneling resonance
    energies show strong non-linear dependence on $H$, with hysteresis and jumps at the
    switching field $|H_{\textrm{sw}}|$.
    For fields beyond the switching field, the levels can
    exhibit non-monotonic variations as a function of $H$, with each level
    behaving differently. For a more detailed discussion of the magnetic field dependence
    of the tunneling energies see Refs.~\onlinecite{mandar,prl_kleff}.
    \\
    (P3): In the large-field regime ($|H| \gg |H_{\textrm{sw}}| $), the resonances
    depend  roughly \emph{linearly} on $H$, with $H$-slopes that  almost all
    have the \emph{same sign} for a given grain, \textit{i.e.}\   slopes of opposite
      signs due to Zeeman-shifting of spin-up and spin-down levels
    \cite{RBT,drago} are not observed.
    \\
   (P4): Measurements
   on  a gated  device~\cite{mandar} showed that
    the observed resonances
    correspond predominantly to the tunneling of minority electrons.
    \\
    (P5): Measurements
   on  a gated  device~\cite{mandar} also showed  that
    some of the resonances must be due to nonequilibrium
    excitations, since some resonances disappear when the Coulomb
    blockade threshold for the onset of tunneling (i.e.\ the amount
    of nonequilibrium) is reduced by tuning the gate voltage to
    lie close to a degeneracy point.
\begin{figure}[t]
{\includegraphics[clip,width=0.98\linewidth]{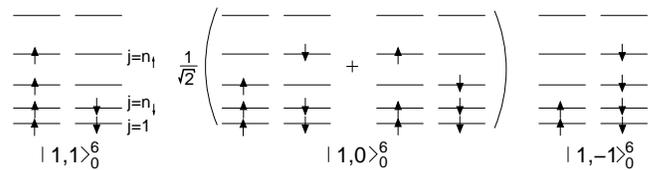}}%
\caption{\label{fig:multi1}The spin wave multiplet $\ket{s,m}_{0}^{N}$
for $s=1$, $N=6$. The position of each level represent
    its  kinetic energy $\varepsilon_j$.}
\end{figure}

\section{\label{sec:model}Model}

In this section we introduce a simple model for ultrasmall
ferromagnetic grains. The  challenge is to describe the
individual quantized electronic excitations of a ferromagnetic
nanoparticle, taking into account the electronic correlations
induced by magnetic interactions and anisotropy forces.

\subsection{\label{sec:hamiltonian}Hamiltonian}
We propose to model a nanoscale magnet with discrete
excitations by the following ``minimal'' Hamiltonian:\cite{prl_kleff,CMcD}
\begin{subequations}
\label{eq:hamiltonian}
\begin{eqnarray}
{\cal H}&=&{\cal H}_{\textrm{C}} + {\cal H}_{0}+
{\cal H}_{\textrm{exch}} +{\cal H}_{\textrm{Zee}} +
{\cal H}_{\textrm{uni}}\; ,
\\
\label{HEC}
{\cal H}_{\textrm{C}} & = &
eV_D{\delta\hat N}+E_{\textrm{C}}{\delta\hat N}^2\; ,
\\
\label{H0}
{\cal H}_0 & = & \sum_{j \sigma} \varepsilon_{j}
      c_{j \sigma}^\dagger c_{j \sigma} \; ,
      \\
\label{Hexch}
{\cal H}_{\textrm{exch}} &=& - {U \over 2} \vec S \cdot \vec S
   \; ,
\\
{\cal H}_{\textrm{Zee}}  & = &
-h S^z \; ,
\\
\label{Huni}
{\cal H}_{\textrm{uni}} &=& -k_N (\vec S \cdot \hat n)^2 /s_0 \, ,
\end{eqnarray}
\end{subequations}
\noindent
with $ h=g_{\text{eff}}\mu_{\textrm{B}}\mu_{0} H$. Here
${\cal H}_{\textrm{C}}$ is the standard Coulomb charging energy
for a nanoparticle with $\delta N$ excess electrons.
$V_D = \left[ Q_0 + \sum_r V_r C_r\right] / C$,
for $r=(L,R,g)$, is the electrostatic potential energy of the island,
with $C\equiv \sum_{r}C_r$. $C_L$, $C_R$ and $C_g$ are
the tunnel junctions connecting the grain to the left or
right lead or the gate electrode, respectively, and
$V_L$, $V_R$ and $V_{\textrm{g}}$ are the voltages of the
left/right leads and the gate electrode.
$Q_0$ is an initial random off-set charge and
$E_{\textrm{C}} = e^2/2C$ is the unit of charging energy.

${\cal H}_0$
describes the kinetic energy of a single band of single-electron
states $|j, \sigma \rangle$, labeled by a discrete index $j$ and a
spin index $\sigma = \text{$(\uparrow, \downarrow)$}$, with the spin
quantization axis chosen in the $z$-direction.  The exchange, Zeeman,
and anisotropy terms, ${\cal H}_{\textrm{exch}}$, ${\cal H}_{\textrm{Zee}}$ and
${\cal H}_{\textrm{uni}}$, are functions of the
total spin vector $\vec S = \sum_j \vec S_j$, where $\vec S_j$ is
the spin vector of the electrons in
level $j$,
\begin{eqnarray}
\label{eq:level_j_spin}
\vec S_j = \frac{1}{2} \sum_{\sigma'\sigma} c^\dagger_{j \sigma}
\vec \sigma_{\sigma \sigma'} c_{j \sigma'} \; ,
\end{eqnarray}
($\vec \sigma$ are Pauli matrices).
${\cal H}_{\textrm{exch}}$ is a rotationally
invariant term which models the effects of an exchange field and
forces the system to adopt a non-zero total ground state spin, say
$s_0$.  On account of this term, spins aligned parallel or
antiparallel to $\langle \vec S\rangle $ may be thought of as forming
``majority'' and ``minority'' bands, which effectively rotate rigidly
together with the magnetization direction.  We shall take the mean
level spacings near the
respective Fermi energies,
$d_{\textrm{min}}\approx 1.19\,\textrm{eV}/s_0$ and
    $d_{\textrm{maj}}\approx 4.61\,\textrm{eV}/s_0$,
    and the exchange-splitting of the Fermi energies,
$\Delta_{\textrm{F}} \equiv \varepsilon_{\textrm{F,maj}} -
\varepsilon_{\textrm{F,min}}$ ($\approx 2$~eV for
Co),~\cite{Papa} as characteristic parameters of the model.

${\cal H}_{\textrm{Zee}} $ describes the spin Zeeman energy in an
external magnetic field $\vec{H} = H \hat z$.  Finally, the
uniaxial anisotropy ${\cal H}_{\textrm{uni}}$ is the simplest
non-trivial form of an anisotropy modeling the combined effects
of crystalline, shape and surface anisotropies,
etc.~\cite{anisotropy} $\hat n$ is the unit vector in the
easy-axis direction and $k_N (>0)$ is a volume-independent
constant. $k_N$ can be estimated from the measured switching
field using $k_N\approx \mu_0\mu_BH_{\textrm{sw}}$,
which yields $k_N\approx 0.01$meV [see Ref.~14 in
Ref.~\onlinecite{prl_kleff}]. For completeness we note that our
analysis in a previous paper [Ref.~\onlinecite{prl_kleff}, see
also Ref.~\onlinecite{mandar}] showed that the anisotropy
constant $k_N$ undergoes fluctuations, \textit {i.e.}\  varies
between different electronic states within a ferromagnetic
nanoparticle. It was shown that these fluctuations have
significant consequences for the magnetic field dependence of the
resonance energies. However, in the present paper we shall
focus only on understanding the characteristic spacings
between resonances observed in DGR's experiments, and not on
their magnetic field dependence; we will therefore neglect
fluctuations in $k_N$ throughout this paper.

\subsection{\label{sec:basis_states}Basis states}
Let us for the moment set the anisotropy strength
${k_N}=0$, and use the
eigenstates of ${\cal H} ({k_N}=0)$ to construct a convenient set of
``bare'' basis states.  Since the Hamiltonian then commutes with the total
spin, these states can be
grouped into spin multiplets that are labeled by their $\vec S \cdot
\vec S$ and $S^z$ eigenvalues, say $s(s+1)$ and $m$.
For example, the
bare ground state of ${\cal H}({k_N}=0)$ for given $N$, $s$ and $h
\,(>0)$, say $\ket{s,s}^N_0$, can be written explicitly as a member of the
following multiplet
of normalized states, $ \ket{s,m}^N_0$ [illustrated in
Fig.~\ref{fig:multi1}]:
\begin{eqnarray}
\label{highestsstate}
\ket{s,s}^{N}_{0}
& \equiv & \prod_{j=1}^{n_{\uparrow}}c^{\dagger}_{
j\uparrow} \; \prod_{j=1}^{n_{\downarrow}}c^{\dagger}_{
j\downarrow} \ket{\text{vac}}.
\\
\label{sm}
\ket{s,m}^{N}_{0} & \equiv & \sqrt{\frac{(s+m)!}{(2s)!(s-m)!}}
\;(S_{-})^{(s-m)}\ket{s,s}^{N}_{0}\; ,
\end{eqnarray}
Here $n_{\uparrow/\downarrow} = N/2 \pm s$, and
$S_{-}=\sum_{j}c_{j\downarrow}^{\dagger}c_{j\uparrow}$ is the
spin-lowering operator.  Within this model, the energy difference
between the spin multiplet $|s,m\rangle$ and all other
states that are constructed from the same single-electron levels
is at least of order
$\varepsilon_{\textrm{F,maj}}-\varepsilon_{\textrm{F,min}}$, a
very large value ($\simeq 2$~eV for Co).\cite{Papa}

The inclusion of a non-zero anisotropy term, ${\cal H}_{\textrm{uni}}$,
will cause the true low-energy eigenstates, $\ket{s,m}^N$, to be
linear superpositions of the bare states in the multiplet
$\ket{s,m}^N_0$.
We choose labels such that $\ket{s,m}^N
\to \ket{s,m}^N_0$ as ${k_N}/h \to 0^+$. We shall call the states
$\ket{s,m}^N$ the \emph{spin wave multiplet}, since each can be viewed
as a homogeneous spin wave.

\subsection{\label{sec:eigenenergies}Eigenenergies}
In the absence of anisotropies (${\cal H}_{\textrm{uni}} =
0$), it is possible to write down explicit expressions for the
low-lying excitation energies of our model.  Let us denote the
excitation energy of a state $\ket{s,m}_0^N$
 relative to the ground state $\ket{s_0,m_0}_0^{N_0}$
(for a gated device) by~\cite{jvdralph}
\begin{eqnarray}
  \label{eq:defineENS}
   \delta E( \delta s,\delta m,\delta N) +  e V_D \delta N \; ,
\end{eqnarray}
with $\delta s=s-s_0$, $\delta m=m-m_0$ and $\delta N=N-N_0$. It
is straightforward to show that the voltage-independent
contribution to Eq.~(\ref{eq:defineENS}) has the
following form, first written down by Canali and
MacDonald:\cite{canali}
\begin{widetext}
\begin{eqnarray}
  \label{eq:excitation-energy}
  \delta E(\delta s,\delta m, \delta N) & = &
(\delta N)^2 \left[ \frac{d_{\textrm{maj}}}{8} +  \frac{d_{\textrm{min}}}{8} + E_{\textrm{C}} \right]
+ \delta N \left[ \bar \varepsilon_{\textrm{F}} +
\frac{d_{\textrm{maj}}}{4} +  \frac{d_{\textrm{min}}}{4}\right]
 + {(\delta s)^2 \over 2} \left[d_{\textrm{maj}}
 + d_{\textrm{min}} - U \right]\nonumber \\*
& &+ \delta s \left[\Delta_{\textrm{F}} - U(s_0 + 1/2) + \frac{d_{\textrm{maj}}}{2}
- \frac{d_{\textrm{min}}}{2} \right] + \delta s \, \delta N \left[
\frac{d_{\textrm{maj}}}{2} - \frac{d_{\textrm{min}}}{2} \right]-h\delta m \; .
\end{eqnarray}
\end{widetext}
Here we introduced the average of majority- and minority-band
Fermi energies ${\bar
\varepsilon}_{\textrm{F}}=(\varepsilon_{{\textrm{F,maj}}}+\varepsilon_{{\textrm{F,min}}})/2$.
The  stability of the ground state spin $s_0$, \textit{i.e.}\ the
requirement that $\delta E(\delta s,\delta m, \delta N) > 0$,
implies \cite{canali} the relation $\Delta_{\textrm{F}} = U (s_0
+ 1/2) + d_0$, where $d_0 (\sim 1/s_0)$ is a small,
grain-dependent energy satisfying
\begin{eqnarray}
\label{eq:inequality}
- ( d_{\textrm{maj}} - U/2)+h < d_0 < d_{\textrm{min}}
- U/2+h\;.
\end{eqnarray}
Hence, the magnitude of $U$ may  be estimated as $U \simeq
\Delta_{\textrm{F}}/s_0 \simeq 2 \textrm{eV}/s_0$. Note that the
ground state spin $s_0$ can be changed by a sufficiently large
change in applied magnetic field. However, the range over which
the applied magnetic field has to be swept between two successive
changes in ground state spin is of order $\delta h
\simeq(d_{\textrm{maj}}+d_{\textrm{min}}-U)/2$, which is large
in nanomagnets ($\mu_0 \delta H \gtrsim 25$~T for Co particles
with diameters $\lesssim 4$~nm). Therefore, for a given value of
$N$, we shall, as long as we neglect nonequilibrium effects,
consider only the ``ground state'' spin value $s_0$.

\section{\label{sec:equilibrium}Equilibrium transitions}

To construct the tunneling spectrum associated with, say, adding
an electron to the grain, we must in principle calculate the
excitation energies $\Delta E^{+}_{fi}$ for all the allowed
transitions, \textit{i.e.}\ those for which the tunneling matrix
element
\begin{eqnarray}
\label{eq:matrixelement}
M_{fi}^{j\sigma} \equiv {}^{N+1} \!  \bra{f} c^\dagger_{j \sigma} \ket{i}^N
\end{eqnarray}
is nonzero.~\cite{jvdralph}  For now, we shall neglect
nonequilibrium effects and thus consider only those tunneling
processes for which the initial state corresponds to the grain's
ground state, \textit{i.e.}\ $\ket{i}^{N}=\ket{s_0,s_0}^{N}$. In
particular, we shall focus on two different types of equilibrium
transitions: (A) transitions involving single-particle excitations
whose resonance spacing, estimated from the electron density of
states, is found to be much larger than observed in DGR's
experiments and (B) transitions between different spin wave
states, of which  only two transitions are found to have
significant weight, leaving unexplained the large number of
resonances seen in experiments.

\begin{table*}
\caption{\label{tab1}
Matrix elements $M_{fi}^{j \sigma}$  and excitation energies
$\Delta E_{fi}^{+}$
of those final states $\ket{f}^{N+1}=\ket{s_{f},m_{f}}_{0}^{N+1}$
that can be reached from the initial state
$\ket{i}^{N}=\ket{s_0,s_0}_{0}^{N}$ by adding a spin-$\sigma$
electron to level $ j_{1}=n_{\uparrow}+1$ or
$j_{2}=n_{\downarrow}+1$.}
\begin{ruledtabular}
\begin{tabular}{cccc}
($j \sigma$) & $s_{f},m_{f} $
& $M_{fi}^{j\sigma}$
& $\Delta E^{+}_{fi}$ \\ \hline
($j_1 \uparrow$)& $s_0+1/2,s_0+1/2$& $1$ & $E_{\textrm{C}}+\varepsilon_{\textrm{F,maj}}+ d_{\textrm
{maj}}-(U/2)(s_0+3/4)-h/2$ \\
($j_1 \downarrow$)& $s_0+1/2,s_0-1/2$& $1/\sqrt{2s_0+1}$ &
$E_{\textrm{C}}+\varepsilon_{\textrm{F,maj}}+ d_{\textrm
{maj}}-(U/2)(s_0+3/4)+h/2$ \\
($j_2 \downarrow$)& $s_0-1/2,s_0-1/2$& $1$ & $E_{\textrm{C}}+\varepsilon_{\textrm{F,min}}+ d_{\textrm
{min}}+(U/2)(s_0+1/4)+h/2$  \\
\end{tabular}
\end{ruledtabular}
\end{table*}

\subsection{\label{sec:equilibrium_single_particle}Single particle excitations}
Apart from the multiplets  $\ket{s,m}_0^N$ discussed in
Sec.~\ref{sec:basis_states}, higher energy multiplets can be
built by creating additional single-particle excitations,
\textit{e.g.}\ by starting from the bare multiplet constructed by
applying the spin-lowering
operator to the state $ c^\dag_{j^\prime \uparrow}
c_{j^{\phantom{\prime}} \uparrow}^{\phantom {\dag}} \ket{s,s}_0^N$, with
$n_\downarrow < j \le n_\uparrow$ and $j^\prime > n_\uparrow$.
However, their eigenenergies lie higher than those of the spin wave
multiplet $ \ket{s,m}_0^N $ by an amount of order
$\varepsilon_{j^\prime} - \varepsilon_j$; this is at least of order
the single-electron level spacing, $d_{\textrm{maj}}=4.61\,\textrm{eV}/s_0$
(respectively $d_{\textrm{min}}=1.19\,\textrm{eV}/s_0$),
i.e.\ rather large compared to $d_{\textrm{obs}}$ [cf.\ (P1)]; thus
the mechanism causing the observed abundance of low-energy
excitations, whatever it is, cannot involve
only purely single-particle excitations, but most also involve
spin excitations.

\subsection{\label{sec:equilibrium_spin_waves}Spin wave excitations}

 We shall now study transitions between two multiplets of
spin wave states, with  initial states
$\ket{i}^N=\ket{s_i,m_i}^N$ and final states
$\ket{f}^{N+1}=\ket{s_f,m_f}^{N+1}$. Consider first the
large-field regime $h \gg h_{\textrm{sw}}$, [where
$h_{\textrm{sw}}=g_{\textrm{eff}} \mu_B \mu_0 H_{\textrm{sw}}$].
Since here ${\cal H}_{\textrm{Zee}}$ dominates over ${\cal
H}_{\textrm{uni}}$, we may set ${k_N}=0$ and construct the matrix
elements $M_{fi}^{j\sigma}$ using the \emph{bare} spin wave
multiplets $\ket{s_i,m_i}_0^N$ and $\ket{s_f,m_f}_0^{N+1}$. The
condition $M_{fi}^{j\sigma} \neq 0$ then implies the {\em spin
selection
    rules}
$|s_{f}-s_{i}|=1/2$ and $ |m_{f}-m_{i}|=1/2$.

Among all possible final states
$\ket{f}$ satisfying these selection rules, Table~\ref{tab1} lists those three
which can be reached by adding an electron to the \emph{lowest available}
levels of $\ket{s_0,s_0}^N_0$, namely $j_1 = n_\uparrow + 1$ and $j_2 =
n_\downarrow + 1$: A spin$\uparrow$-electron can be added only to level $j_1$
(Table~\ref{tab1}, row 1), whereas a $\downarrow$-electron can be added to
either level $j_1$ (row 2) or $j_2$ (row 3).

The excitation energies $\Delta
E_{fi}^+$ (Table~\ref{tab1}, column 4) of the ($j_1 \! \uparrow$) and ($j_1 \!
\downarrow$) transitions are degenerate at $h=0$ and Zeeman-split as a
function of $h$, but, in accord with (P3),
this splitting will not be observable: the weight of the ($j_1 \!
\downarrow$) transition is smaller than that of the ($j_1 \!  \uparrow$)
transition (Table~\ref{tab1}, column 3) by a Clebsch-Gordan coefficient of
order $1/(2s_0)$  which is negligibly small for large-$s_0$
grains.~\cite{gueron}
The $(j_1 \!  \uparrow)$ and $(j_2 \!  \downarrow)$ transitions both
have large,
comparable weights, and would produce resonances with large-$h$ slopes of
opposite signs. Depending on whether
the difference in their excitation energies
  (Table~\ref{tab1}, column 4)
is close to or far from 0
(it is at most of order $d_{\textrm{maj}}-U/2$, \textit{i.e.}\
$\lesssim 3$meV for a $4$-nm
-diameter Co particle\cite{Papa}),
either both or only one
of the $(j_1 \!  \uparrow)$ and $(j_2 \!  \downarrow)$
transitions would be observable in the regime of lowest excitation
energies (say $\lesssim 0.5$meV).
However, in an equilibrium tunneling scenario, this leaves unexplained the
large observed density of tunneling resonances (P1), since, apart from
these two
transitions, there are no others with significant weight \emph{and} excitation
energies less than $d_{\textrm{min}}$ ($d_{\textrm{maj}}$).

\begin{figure}[t]
{\includegraphics[width=0.98\linewidth]{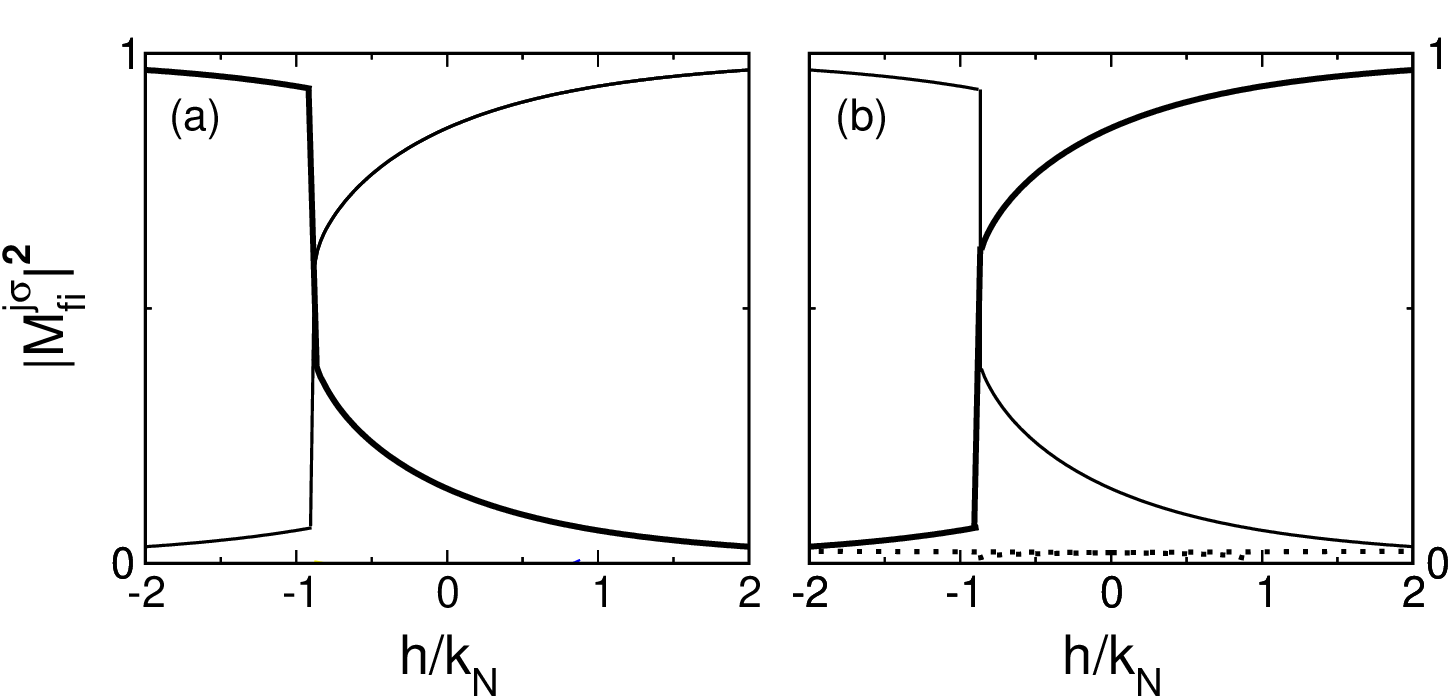}}
\caption{\label{fig:overlaps}Matrix
elements $M_{fi}^{j \sigma}$  for the transitions $\ket{s_i,s_i}^N
\to \ket{s_f,m_f}^{N+1}$ ($s_i=20$), calculated for ${\cal H}_{\textrm
  {uni}}+{\cal H}_{\textrm{Zee}}$ and
 plotted as functions of $h/k_N$ sweeping positive to negative, for the
following final states:
(a):
$\ket{s_i-{1 \over 2},s_i-{1 \over 2}}^{N+1}$;
(b): $ \ket{s_i+{1 \over
2},s_i+{1 \over 2}}^{N+1}$ (solid lines) and $ \ket{s_i+{1 \over 2},s_i-{1
\over 2}}^{N+1}$ (dotted lines).  Compared to these, all other final
states have negligible matrix elements. The  solid lines are
thick for $M^{j \uparrow}_{fi}$ and thin for $M^{j \downarrow}_{fi}$
(both contribute to the \emph{same} transition $\ket{i} \to
\ket{f}$).}
\end{figure}

Next we argue that this problem persists also for the case with anisotropy,
where ${k_N} \neq 0$.  \emph{A priori} one might have expected to see more
low-lying excitations then, since the selection rule $|m_{f}-m_{i}|=1/2 $ no
longer applies if the matrix elements $M_{fi}^{j\sigma}$ are formed using the
exact spin wave multiplets $\ket{s_0,m_i}^N$ and $\ket{s_0\pm 1/2,m_f}^{N+1}$,
which are not $S^z$ eigenstates.  We have therefore numerically
diagonalized ${\cal
    H}_{\textrm{Zee}} + {\cal H}_{\textrm{uni}}$,~\cite{kleff}
 as function of $h/k_N$, for a few
selected values of $s_0$, to determine the eigenstates $\ket{s_0,m_i}^N$ and
$\ket{s_0\pm 1/2,m_f}^{N+1}$ and calculate the  matrix
elements $M_{fi}^{j\sigma}$
[Fig.~\ref{fig:overlaps}], for both spin-increasing and -decreasing
transitions ($s_f = s_0 \pm 1/2$, $j = n_{\uparrow/\downarrow} + 1$).
We find
that in both cases, the transition probability $\sum_\sigma
|M_{fi}^{j\sigma}|^2$ from $\ket{s_0,s_0}^N$ to $\ket{s_0 \pm
    1/2, s_0 \pm 1/2 - n}^{N+1}$ is very much larger for $n = 0$
  than for \emph{any} other $n \neq 0$ state. This is
the same trend as that found in Table~\ref{tab1}.   Thus, even though
${\cal H}_{\textrm{uni}}$ causes violations of one of the spin selection rules,
the extra transitions have too little weight to explain the large density
of low-energy excitations that is observed (P1).~\cite{canali-mcd}

Apart from the fact that only two of the above discussed transitions
have significant weight there are two additional important considerations
which lead us to conclude
that the abundance of resonances seen in experiments cannot be
explained  by equilibrium spin wave transitions alone:
 (i) Firstly, the resonances associated with final states $\ket{s_f,m_f}$
 that differ only in $m_f$ would  have a spacing of order $k_N$
[$\approx 0.01$ meV], \textit{i.e.}\  much smaller than
 the observed resonance
spacing. (ii) Secondly, for high magnetic fields
these resonances would
exhibit a systematic increase in the magnitude of their
slopes (which are $\propto |s_i-m_f|$)
 that was not observed in experiment.
We therefore assert that
the large density of resonances cannot be explained by equilibrium
transitions alone; we will explore nonequilibrium effects below.

\section{\label{sec:nonequilibrium}Nonequilibrium transitions}

Since the large density of resonances (P1)
cannot be explained by equilibrium transitions [neither single
particle excitations (Sec.~\ref{sec:equilibrium_single_particle})
nor spin wave excitations (Sec.~\ref{sec:equilibrium_spin_waves})],
we shall in this section  explore nonequilibrium effects.
Our conclusion will be that a combination of nonequilibrium spin and
single-particle excitations produces a much denser spectrum of
tunneling states than expected within an independent-electron model.
For a spin of $s_0 \simeq 1000$, this nonequilibrium
scenario gives a resonance spacing of $\simeq 0.2$meV for the
spacing of resonances due to the tunneling of minority electrons,
in accord with the observed resonance spacing.

Nonequilibrium spin accumulation had of course already been
studied previously in the context of single-electron transistors with
ferromagnetic components,\cite{brataas,fert1}
and spin accumulation for nanograins
with discrete energy levels was first analysed
by Barn\'as \emph{et al.}~\cite{barnas-martinek}
 However, these analyses
all employed a single-particle description in which
all states that were considered were simple
Slater determinants of single-particle states.
Within our present model, we have to
go beyond this simple picture by considering the
true many-body eigenstates of the Hamiltonian,
which are in general \emph{linear combinations} of
Slater determinants.

After explaining the general idea of nonequilibrium processes in
ferromagnetic grains in Sec.~\ref{sec:general_master_equation}, we
shall describe  different nonequilibrium scenarios.
For each, we calculate the corresponding theoretical tunnel
spectra and compare resonance spacings and the number of
resonances with DGR's measurements.  We show in
Sec.~\ref{sec:nonequilibrium_spin_ground_states}
 that nonequilibrium spin excitations lead to resonance spacings
as observed in measurements and in
Sec.~\ref{sec:nonequilibrium_spin_wave} that Zeeman splittings
are suppressed for large spin $s_0$. A combination of spin- and
single particle excitations
(Sec.~\ref{sec:nonequilibrium_single_particle}) significantly
enhances the number of tunneling resonances achievable
for a given setting of the gate voltage, making it possible to
explain the large number of resonances observed by DGR
even for the case of a small Coulomb blockade threshold
(\textit{i.e.}\ weak nonequilibrium).

\subsection{\label{sec:general_master_equation}General master equation}

\begin{figure}[t]
{\includegraphics[clip,width=0.6\linewidth]{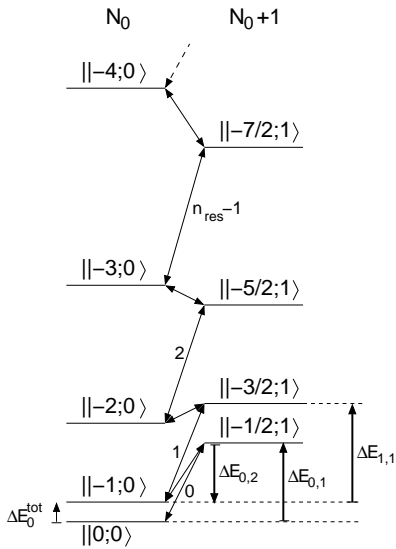}}
\caption{\label{fig:spin_ground_ladder}Illustration of a
nonequilibrium scenario involving only spin ground states,
$||\delta s;\delta N\rangle\equiv\ket{s_0+\delta s,s_0+\delta s}^{N_0+\delta N}$,
for  the case that the first electron that tunnels
is a minority electron ($\alpha =-1$) that enters the grain ($p=+1$).
Here $\delta s$ and $\delta N$ characterize
the spin with respect to the  overall spin ground state: $\delta s=s-s_0$
and $\delta N=N-N_0$ (see Secs.~\ref{sec:eigenenergies}
and~\ref{sec:nonequilibrium_spin_ground_states}).
The vertical arrows indicate the energy differences
$\Delta E_{n,1}^{\alpha p}$, $\Delta E_{n,2}^{\alpha p}$ and
$\Delta E_0^{\textrm{tot}}$. Charging transitions are
numbered as $n=0$, $1$, $2\cdots$.}
\end{figure}

 In general, $N$-electron states other than the
ground state can be populated during the process of current flow, and
this may affect the experimental tunneling spectrum.~\cite{agam,jvdralph}
A simple scenario
is illustrated in Fig.~\ref{fig:spin_ground_ladder}.
Even if a first
tunneling event causes a ``charging'' transition from the $N$-electron
ground state $| {\gamma_{\textrm{g}}} \rangle^N$ to the $(N \pm 1)$-electron ground
state $|{\gamma_{\textrm{g}}} \rangle^{N \pm 1}$, it may be energetically possible
for the subsequent ``discharging'' tunneling transition to return the
particle to an excited $N$-electron state $|\gamma_{\textrm{e}} \rangle^N$ instead
of $|{\gamma_{\textrm{g}}} \rangle^N$, provided the applied voltage is sufficiently
large, $eV \gtrsim E_{\textrm{e}}^N - E_{\textrm{g}}^N$.
Likewise, further charging
and discharging transitions may allow any of a large ensemble of
states to be occupied at higher and higher levels of an energy ladder,
terminating only when an energy-increasing transition requires more
energy than the applied voltage provides.  As the voltage is
increased, the total current (or conductance) may increase stepwise
(or show peaks) when thresholds are crossed to allow higher-energy transitions
up the nonequilibrium ladder, thereby changing the occupation
probabilities of the ensemble of nonequilibrium states and opening new
tunneling channels.

Let $\{\gamma\}$ be the set of all states involved in a
nonequilibrium ladder of excitations, \textit{i.e.}\  the set of
all discrete states with a nonzero occupation probability
$P(\gamma)$ for a given bias voltage, gate voltage,
temperature and magnetic field.
To find the occupation probability $P(\gamma)$
for all the  states $\ket{\gamma}$ of the ladder
one has to solve a normalization condition
$\sum_{\gamma^{\prime}}P({\gamma^{\prime}})=1$
and a stationary master equation
of the form\cite{jvdralph}

\begin{eqnarray}
\label{eq:master_equation}
0=\sum_{{\gamma^{\prime}}\not= {\gamma}}
 \Bigl\{\Sigma_{{\gamma}{\gamma^{\prime}}}\Bigr. P({\gamma^{\prime}})
 &-& \Sigma_{{\gamma^{\prime}}{\gamma}} P({\gamma})
\\* \nonumber
+\Sigma_{{\gamma}{\gamma^{\prime}}}^{\textrm{el}}P({\gamma^{\prime}})
 &-& \Sigma_{{\gamma^{\prime}}{\gamma}}^{\textrm{el}} P({\gamma})
\\*
+\Sigma_{{\gamma}{\gamma^{\prime}}}^{\textrm{sf}}P({\gamma^{\prime}})
 &-& \Sigma_{{\gamma^{\prime}}{\gamma}}^{\textrm{sf}}
 \Bigl.P({\gamma})\Bigr\}\;,\nonumber
\end{eqnarray}
for each ${\gamma}$. The first (second) term in Eq.
(\ref{eq:master_equation}) describes the rate at which the
probability of a given configuration increases (decreases) due to
electrons tunneling onto or off the grain, the remaining
terms are associated with  electronic relaxation and
spin-flip relaxation on the grain, respectively.

$\Sigma_{{\gamma}{\gamma^{\prime}}}$ is the total
tunneling-induced transition rate from initial state
$\ket{\gamma^{\prime}}$ to final state $\ket{\gamma}$.
Considering sequential tunneling only, it has the form
\begin{eqnarray}
\Sigma_{{\gamma}{\gamma^{\prime}}}
=\sum_{r=L,R}\sum_{p=\pm}
\Sigma^{rp}_{{\gamma}{\gamma^{\prime}}}\; ,
\end{eqnarray}
where $\Sigma^{r+}_{{\gamma}{\gamma^{\prime}}}$
($\Sigma^{r-}_{{\gamma}{\gamma^{\prime}}} $)
involves the coherent transfer of an electron onto
(from) the grain from (onto) lead $r$ and is
given by
\begin{eqnarray*}
\Sigma^{r+}_{{\gamma}{\gamma^{\prime}}}
&=&\Gamma^r \sum_{i\sigma}
|\bra{\gamma}c^{\dagger}_{i\sigma}\ket{{\gamma^{\prime}}}|^2
f(E_{{\gamma}}-E_{{\gamma^{\prime}}}-e\bar{V}_r)\;,
\\
\Sigma^{r-}_{{\gamma}{\gamma^{\prime}}}
&=&\Gamma^r \sum_{i\sigma}|\bra{{\gamma} }c_{i\sigma}\ket{{\gamma^{\prime}}}|^2
f(E_{{\gamma}}-E_{{\gamma^{\prime}}}+e\bar{V}_r)\;.
\end{eqnarray*}
Here $f(E)=1/(e^{E/k_B T}+1)$ is the Fermi function and
$e\bar{V}_r$ is the electrostatic potential energy difference
between lead  $r$ and the grain,
\begin{eqnarray}
  \label{eq:definebarV}
  e \bar V_r \equiv eV_r  - e V_D \; .
\end{eqnarray}

The third and fourth terms in Eq. (\ref{eq:master_equation})
describe electronic (spin-conserving) relaxation processes inside
the grain. For simplicity, we shall only consider electronic
relaxation between energetically ``neighboring'' single particle levels,
i.e.\ we take
\begin{eqnarray}
\Sigma_{{\gamma}{\gamma^{\prime}}}^{\textrm{el}}&=&\Gamma^{\textrm{el}}
\sum_{i\sigma}
|\bra{{\gamma} }
c^{\dagger}_{(i-1)\sigma}c_{i\sigma}\ket{{\gamma^{\prime}}}|^2\;.
\end{eqnarray}
(Generalizations of this assumption are straightforward,
though cumbersome.)
 The last two terms of Eq.~(\ref{eq:master_equation}) describe
the rate at which the probability of a given distribution
increases (decreases) due to spin-flip relaxation in the
ferromagnetic grain. For simplicity we shall assume
all spin-flip relaxation rates to be much smaller than all other
rates, $\Gamma^{\textrm{sf}} < \Gamma^{\textrm{r/l}},
\Gamma^{\textrm{el}}$, and hence take $\Sigma^{\textrm{sf}}=0$
throughout this paper.\cite{rates}) (Again, it is
straightforward to consider generalizations of this case.) Moreover, all
rates $\Gamma$ are assumed to be independent of the specific
single particle level $i\sigma$ involved.

The current trough the grain can then be calculated as~\cite{jvdralph}
\begin{eqnarray}
I_r=e\sum_{{ \gamma}{\gamma^{\prime}}}
\left(\Sigma^{r+}_{{\gamma}{\gamma^{\prime}}}
-\Sigma^{r-}_{{\gamma}{\gamma^{\prime}}}\right) P({\gamma^{\prime}})\,.
\end{eqnarray}

\subsection{
\label{sec:nonequilibrium_spin_ground_states}Spin accumulation}

In a ferromagnetic particle, in addition to the nonequilibrium
occupation of single-electron states discussed previously for
non-magnetic particles,~\cite{agam} nonequilibrium spin
excitations are possible, too, if the spin-flip rate
$\Gamma^{\textrm{sf}}$ is smaller than the tunneling rate
$\Gamma^{\textrm{tun}}$.~\cite{rates} In this case a ladder of
transitions
  will occur between states with
  different total spin $s$, causing each to have a finite occupation
  probability and thus leading to {\em spin accumulation} on the
  grain.~\cite{barnas-martinek,fert1,brataas}

The simplest nontrivial case, namely a  ladder of spin multiplet
ground states $\ket{s,s}$ (see
Fig.~\ref{fig:spin_ground_ladder}) was already briefly discussed
in Ref.~\onlinecite{prl_kleff}. Below we shall discuss this case
in more detail. We  shall calculate the resonance
spacing of steps in the current (or of peaks in the differential
conductance) and the number of resonances for transitions between
spin ground states. We shall characterize spin ground states by
their {\em spin and charge relative to the overall ground state}, \textit{i.e.}\
we shall write 
\begin{equation}
||\delta s;\delta N\rangle\equiv
\ket{s_0+\delta s,s_0+\delta s}^{N_0+\delta N}\; ,
\end{equation}
with $\delta s$ and $\delta N$ as defined in Sec.~\ref{sec:eigenenergies} 
after Eq.~\ref{eq:defineENS}.
We shall find that the resonance spacing agrees
very well with DGR's measurements.

Consider a sequence
of nonequilibrium transitions forming a ``ladder"  ($L^{\alpha p} $)
with ``rung index'' $n = 0, 1, 2, \dots$,
where each rung corresponds to a ``charging'' transition
($L^{\alpha p}_{n,1} $) followed by a ``discharging'' transition"
($L^{\alpha p}_{n,2}$):
\[
  \label{eq:rung-n-transitions}
  ||\alpha p n;0\rangle
  \stackrel{L^{\alpha p}_{n,1}}{\longrightarrow}
  ||\alpha p (n+1/2);p\rangle
  \stackrel{L^{\alpha p}_{n,2}}{\longrightarrow}
  ||\alpha p (n+1);0\rangle \; .
\]
(Above, the notation $||\delta s;\delta N\rangle$ is used.)
The indices $p$ and $\alpha$ are used to
distinguish whether the {\em first} electron that tunnels enters
or leaves the grain, $p=(+1,-1)$, and whether it is a majority or
minority electron,  $\alpha = (\textrm{maj},\textrm{min}) = (+1,
-1)$.

Using Eq.~(\ref{eq:excitation-energy}), the   threshold energy costs
for transitions $L^{\alpha p}_{n,1} $ and $L^{\alpha p}_{n,2}$
can be calculated to be:\cite{no_hanis_hzee}
\begin{subequations}
\label{eq:rung-n}
\begin{eqnarray}
  \label{eq:rung-n-E1}
  \Delta E^{\alpha p}_{n,1} & = &
E(\alpha p (n+1/2);p) - E (\alpha p n;0)
\\*
  & = &
  \Delta E^{\alpha p}_{0,1}  +
  n \left[d_\alpha - U/2 \right] \; ,\nonumber
\\
\label{eq:rung-n-E2}
\Delta E^{\alpha p}_{n,2} & = &
E(\alpha p (n+1);0) - E(\alpha p (n+1/2);p)
\quad \phantom{.}
\\*
  & = &
  \Delta E^{\alpha p}_{0,2}  +
  n \left[d_{\bar \alpha} - U/2 \right] \; ,\nonumber
\end{eqnarray}
\end{subequations}
with
\begin{eqnarray*}
 \Delta E^{\alpha p}_{0,1} & = &
d_\alpha (1+p)/2 + E_{\textrm{C}} +
p(\bar \varepsilon_{\textrm{F}} + \alpha d_0 /2)
- U/8  \; , \\
\Delta E^{\alpha p}_{0,2} & = &
d_{\bar \alpha} (1-p)/2
- E_{\textrm{C}} - p(\bar \varepsilon_{\textrm{F}} - \alpha d_0 /2)
-  3\, U/8  \; .
\end{eqnarray*}
(Above, the notation $\bar \alpha$ means $\overline{\textrm{maj}}
= \textrm{min}$ and $\overline{\textrm{min}} = \textrm{maj}$.)
Note that the total energy cost for the
combined transitions $L^{\alpha p}_{0,1}$ and $L^{\alpha
p}_{0,2}$, namely
\begin{subequations}
\begin{eqnarray}
  \label{eq:E0tot}
  \Delta E^{\alpha p}_{0,\textrm{tot}} &\equiv &
\Delta E^{\alpha p}_{0, 1}  + \Delta E^{\alpha p}_{0, 2} \\* & =
&d_\alpha \frac{1 + p}{2} + d_{\bar \alpha}
\frac{1 - p}{2} + p \alpha d_0 - \frac{U}{2} \;   \,  ,
\qquad\phantom{.} 
\end{eqnarray}
\end{subequations}
is always $\ge 0$; this follows intuitively
from the fact that $  \Delta E^{\alpha p}_{0,\textrm{tot}}$
is the excitation energy between the overall ground
state $||\delta s;\delta N\rangle =||0;0\rangle$ and the adjacent-spin ground
state $||\alpha p;0\rangle$, and more formally
from condition (\ref{eq:inequality}) on $d_0$.

Assuming that the peaks in the conductance are due to
successive charging transitions becoming accessible as the bias
voltage is increased, the resonance spacing  for the ladder
$L^{\alpha p}$ can readily be calculated using
Eqs.~(\ref{eq:rung-n}):
\begin{subequations}
\begin{eqnarray}  \label{eq:define-resonance-spacing}
\delta E^{\alpha p}_{\textrm{res}} &\equiv & \Delta
E_{n+1,1}^{\alpha p}-\Delta E_{n,1}^{\alpha p}
\\ &=&  d_\alpha - U/2 \; . \label{eq:spacing}
\end{eqnarray}
\end{subequations}
This result, which evidently depends only on whether the
charging transition involves the tunneling of a majority or
minority electron, $\alpha = (\textrm{maj}, \textrm{min})$,
can be intuitively understood as follows: The resonance spacing,
$\delta E^{\alpha p}_{\textrm{res}}$, is a difference of energy
differences, \textit{i.e.}\ a type of (discrete) second derivative of the
total energy. The contribution $ d_\alpha$ reflects the
discrete second derivative w.r.t.\ the quasiparticle number of the
energy involved in creating, using only $\alpha$-electrons, $n$
particle-like or hole-like excitations  relative to the overall
ground state, and the term $-U/2$ reflects the discrete second
derivative w.r.t.\ the spin of the exchange energy. The partial
cancellation between $ d_\alpha$ and $-U/2$ in
Eq.~(\ref{eq:spacing}) reflects the opposite signs of the kinetic
and exchange energies in the Hamiltonian (\ref{eq:hamiltonian}),
and is thus very generic. Very significantly, since $U/2$ and
$ d_{\textrm{min}}$ (but not $ d_{\textrm{maj}}$) happen to
be almost equal, this partial
cancellation of two rather large energies produces a much smaller
energy,
\begin{subequations}
\label{eq:spacing-estimates}
\begin{eqnarray}
 \delta E^{\textrm{maj}, p}_{\textrm{res}} &\simeq & (4.6 - 1.0
)\textrm{eV}/ s_0 \, = \, \textrm{3.6 eV}/s_0
 \; ,
 \label{eq:spacing-estimates-maj}
\\ \label{eq:spacing-estimates-min}
 \delta E^{\textrm{min}, p}_{\textrm{res}} & \simeq & (1.2 - 1.0
)\textrm{eV}/ s_0  \, = \, \textrm{0.2 eV}/s_0
 \; ,
\end{eqnarray}
\end{subequations}
where we used the parameter estimates given in
Secs.~\ref{sec:experimental_results} and \ref{sec:model}.~\cite{negativ_spacing}
This brings us to one of the main conclusions of this
paper: \emph{the small resonance spacing of} 0.2 meV
\emph{observed by DGR is consistent with prediction}
(\ref{eq:spacing-estimates-min}) \emph{for minority-electron
charging events} if the ground state spin is assumed to be about
$s_0 \simeq 1000$, which is within the estimated size range of
DGR's grains. Satisfactorily, the conclusion that minority
electrons dominate the charging transitions, which was reached
independently by Canali and MacDonald too,\cite{canali} has
recently been confirmed experimentally for DRG's gated device, as
has been the conclusion that nonequilibrium physics is involved
[cf.\ points (P4) and (P5) of
Sec.~\ref{sec:experimental_results}].

We shall therefore henceforth consider only the case that
charging transitions are due to minority electrons, \textit{i.e.}\ we take
$\alpha = \textrm{min}$ (but for notational brevity will
sometimes still use the index $\alpha$ instead of
``$\textrm{min}$''). The conductance will then show a limited
number, say $n_{\textrm{res}}^{\textrm{ch}}$, of resonances due to
\emph{charging} transitions, with a rather small spacing of $
d_{\textrm{min}}- U/2$, followed by an unlimited number  of
resonances due to \emph{discharging} transitions, with a much
larger spacing of $ d_{\textrm{maj}} - U/2$. This can be seen
as follows: let us consider for definiteness  a circuit with $V_L
= - V_R = V/2$, and suppose that $eV > 0$, so that electrons flow
from left to right through the ferromagnetic grain. Then
transition $L^{\alpha p}_{n,1}$ involves
tunneling across the left
junction if $p = 1$, or the right junction
if $p= -1$, [and likewise $L^{\alpha p}_{n,2}$ involves
tunneling across
junction $(R,L)$ for $\bar p = (-1, 1)$]. The voltage thresholds,
say $e V^{\alpha p}_{n,1}$ (or $e V^{\alpha
  p}_{n,2}$), needed to overcome the energy cost $\Delta
E^{\alpha p}_{n,1}$ (or $\Delta E^{\alpha p}_{n,2}$) in order for
the charging transition $L^{\alpha p}_{n,1}$ (or the discharging
transition $L^{\alpha p}_{n,2}$) to occur, are determined by the
conditions:
\begin{subequations}
   \label{eq:V-thresholds}
 \begin{eqnarray}
   p \, e \bar V_p & \ge & \Delta E^{\alpha p}_{n,1} \; ,
\\
   \bar  p \, e \bar V_{\bar p} & \ge  & \Delta E^{\alpha p}_{n,2}
\; .
\end{eqnarray}
\end{subequations}
Here $e \bar V_p$, the electrostatic potential energy
difference between lead $p$ and the grain, is [from
Eq.~(\ref{eq:definebarV})] related to the actual applied voltage
by
\begin{eqnarray}  \label{eq:ebarV}
 e \bar V_p = p \, eV B_p - e \tilde V_{\textrm{g}} \; ,
\end{eqnarray}
where $ B_p \equiv (C_{\bar p}+C_g/2)/C$ is a capacitance ratio
which converts applied voltage to energy,\cite{jvdralph} and
$e\tilde V_{\textrm{g}} \equiv
  e(Q_0+V_{\textrm{g}}C_g)/C$ is an offset energy.
  It follows from Eqs.~(\ref{eq:V-thresholds}) and (\ref{eq:ebarV})
that $V^{\alpha p}_{n,1}$ and $V^{\alpha p}_{n,2}$ are given by
\begin{subequations}  \label{eq:Vn12-explicit}
\begin{eqnarray}
B_p \, eV^{\alpha p}_{n,1} & = & \Delta E^{\alpha p}_{n,1} + p \,
e \tilde V_{\textrm{g}} \; ,
\\
B_{\bar p} \, eV^{\alpha p}_{n,2} & = & \Delta E^{\alpha p}_{n,2}
- p \, e \tilde V_{\textrm{g}} \; .
\end{eqnarray}\end{subequations}
In particular, the threshold voltage $V^{\alpha p}_{0,1}$
for the first charging transition, $L^{\alpha p}_{0,1}$,
determines the measured size of the Coulomb-blockade region (in
energy units), say $E_{\textrm{C}}^{\textrm{thresh}}$:
\begin{eqnarray}
  \label{eq:defineECthresh}
  E_{\textrm{C}}^{\textrm{thresh}} \equiv e V_{0,1}^{\alpha p} B_p =
\Delta E^{\alpha p}_{0,1} + p e \tilde V_{\textrm{g}} \; ,
\end{eqnarray}
 This  enables us to eliminate $p \, e \tilde V_{\textrm{g}}$
 from  Eq.~(\ref{eq:Vn12-explicit}) and write it as [using
 Eqs.~(\ref{eq:rung-n}) and (\ref{eq:E0tot})]
 \begin{subequations}  \label{eq:Vn12-very-explicit}
 \begin{eqnarray}
 B_p \, eV^{\alpha p}_{n,1} & = & n ( d_\alpha - U/2) +
E_{\textrm{C}}^{\textrm{thresh}}
 \; ,
\\
B_{\bar p} \, eV^{\alpha p}_{n,2} & = & n ( d_{\bar \alpha} -
U/2) - E_{\textrm{C}}^{\textrm{thresh}}  + \Delta E_{0, \textrm{tot}
}^{\alpha,p}\; . \quad \phantom{.}
\end{eqnarray}
\end{subequations}

\begin{figure}[t]
{\includegraphics[clip,width=0.98\linewidth]{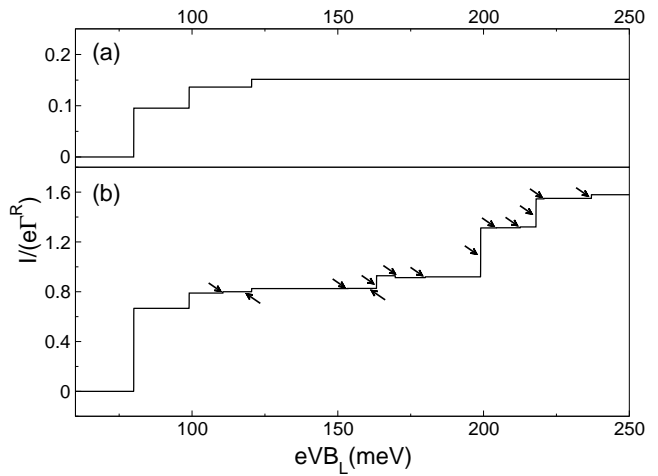}}
\caption{\label{fig:single_particle} Current as function of
$eVB_L$ for the following parameters: $\alpha ={\textrm{min}}$,
$p=+1$, $s_0=10$, ${d}_{\textrm{min}}=1.19\, \textrm{eV}/s_0$,
${d}_{\textrm{maj}}= 4.61\, \textrm{eV}/s_0$, $U=2\,
\textrm{eV}/s_0$, $B_L=0.3$, $B_R=0.7$,
$E_{\textrm{C}}^{\textrm{thresh}}=0.8\,\textrm{eV}/s_0$,
$E_{0,{\textrm{tot}}}^{\alpha p}=0\,\textrm{eV}/s_0$,
$\Gamma^L/\Gamma^R=0.8$, $\Gamma^{\textrm{sf}}=0$, $T=80$ mK.~\cite{smallspin} (a)
Only spin excitations (b) Spin excitations and single-particle
excitations:
 $\Gamma^{\textrm{el}}/\Gamma^R=10^6$.
Arrows mark additional current steps due to combined
spin and single-particle excitations.
No significance should be attached to step heights here,
since they depend on (unknown) tunneling matrix elements,
which for simplicity we took all equal.~\cite{single_particle}
Parameters are chosen according
to Fig. 1 of Ref.~\onlinecite{mandar}.}
\end{figure}
 Now,  whenever
 $E_{\textrm{C}}^{\textrm{thresh}} \gtrsim \frac{1}{2} \Delta E_{0,
 \textrm{tot}}^{\alpha,p}$, \textit{i.e.}\ whenever $E_{\textrm{C}}^{\textrm{thresh}} $
 is not too small, the inequality
\begin{eqnarray}
  \label{eq:V2<V1}
  e V^{\alpha p}_{n,1} > e V^{\alpha p}_{n,2} \, .
\end{eqnarray}
will hold for sufficiently small values of $n$ (at least
for $n= 0$). However, for large enough $n$ it will cease to hold,
since for the case $\alpha = \textrm{min}$ that we are
considering, the ``step size'' $ d_{\textrm{maj}} - U/2$ for $e
V^{\alpha p}_{n,2} $ is much larger than the ``step size'' $
d_{\textrm{min}} - U/2$ for $e V^{\alpha p}_{n,1} $. Thus, the first
few measured conductance resonances will be due to a sequence of
(rather closely-spaced) \emph{charging} transitions, as opposed
to (much more widely-spaced) discharging transitions, because
each time the bias voltage is incremented by $eV^{\alpha
p}_{n,1}$ to make the next charging transition $L^{\alpha
p}_{n,1}$ energetically accessible, this bias voltage increment
is already large enough (namely $> eV^{\alpha p}_{n,2}$) to also
allow the discharging transition $L^{\alpha p}_{n,2}$ to occur.
However, once the inequality Eq.~(\ref{eq:V2<V1}) is violated,
the subsequent discharging transition $L^{\alpha p}_{n,2}$ will
become possible only after the total bias voltage increment
reaches $eV^{\alpha p}_{n,2}$, \textit{i.e.}\ henceforth \emph{discharging}
(instead of charging) transitions will determine the conductance
resonances, which will henceforth be spaced much more widely.

To calculate the total number of closely-spaced resonances
due to charging transitions, $n^{\textrm{ch}}_{\textrm{res}}$, we
must thus determine how large $n$ can become before the condition
Eq.~(\ref{eq:V2<V1}) ceases to hold. Using
Eqs.~(\ref{eq:Vn12-very-explicit}) and $B_p + B_{\bar p} = 1$,
this condition can be rearranged to yield an expression for
$n^{\textrm{ch}}_{\textrm{res}}$, which is found to be given by the
smallest integer larger or equal to
\begin{equation}
\label{eq:n_max}
1+\frac{E_{\textrm{C}}^{\textrm{thresh}}- B_p\Delta
E_{0,{\textrm{tot}}}^{\textrm{min}, p}}{B_p( d_{\textrm{maj}}
- U/2) - B_{\bar p} ( d_{\textrm{min}}-U/2) }\; .
\end{equation}

The prediction that $n^{\textrm{ch}}_{\textrm{res}}$ increases
linearly with $E_{\textrm{C}}^{\textrm{thresh}}$ is in
qualitative agreement with Fig.~2 of
Ref.~\onlinecite{gueron}.~\cite{prl_kleff} However, it is
not quite consistent with more recent data on
Co-grains,~\cite{mandar} where, even when the Coulomb blockade
threshold was very small ($E_{\textrm{C}}^{\textrm{thresh}} \lesssim
1$~meV),  the differential conductance showed many ($> 10$)
peaks, \textit{i.e.}\ many more than Eq. (\ref{eq:n_max}) would predict.
To illustrate this, we have solved Eq.~(\ref{eq:master_equation})
numerically for the discussed transitions   and calculated the
current for the parameters of the model and a threshold charging
energy of $E_{\textrm{C}}^{\textrm{thresh}}=0.8\,
\textrm{eV}/s_0$ as function of $eVB_L$ [see
Fig.~\ref{fig:single_particle}(a)]. The calculated current shows
three steps in the plotted region. Experimental data (see Fig.~1
in Ref.~\onlinecite{mandar}) on the other hand show $\approx 12$
resonances for the parameters used.

To summarize the conclusions of this section: a
nonequilibrium scenario involving only spin ground states, and
assuming charging transitions involving minority electrons,
results in a spacing of resonances, $\delta E_{\textrm{res}}\approx
0.2$ meV for $s_0=1000$, that agrees roughly with the spacing
observed. However, the number of resonances predicted by
Eq.~(\ref{eq:n_max}) is sometimes much smaller than observed,
namely when $E_{\textrm{C}}^{\textrm{thresh}}$ is small (on the
order of $ d_{\textrm{min}}$).

\begin{figure}[t]
{\includegraphics[clip,width=0.98\linewidth]{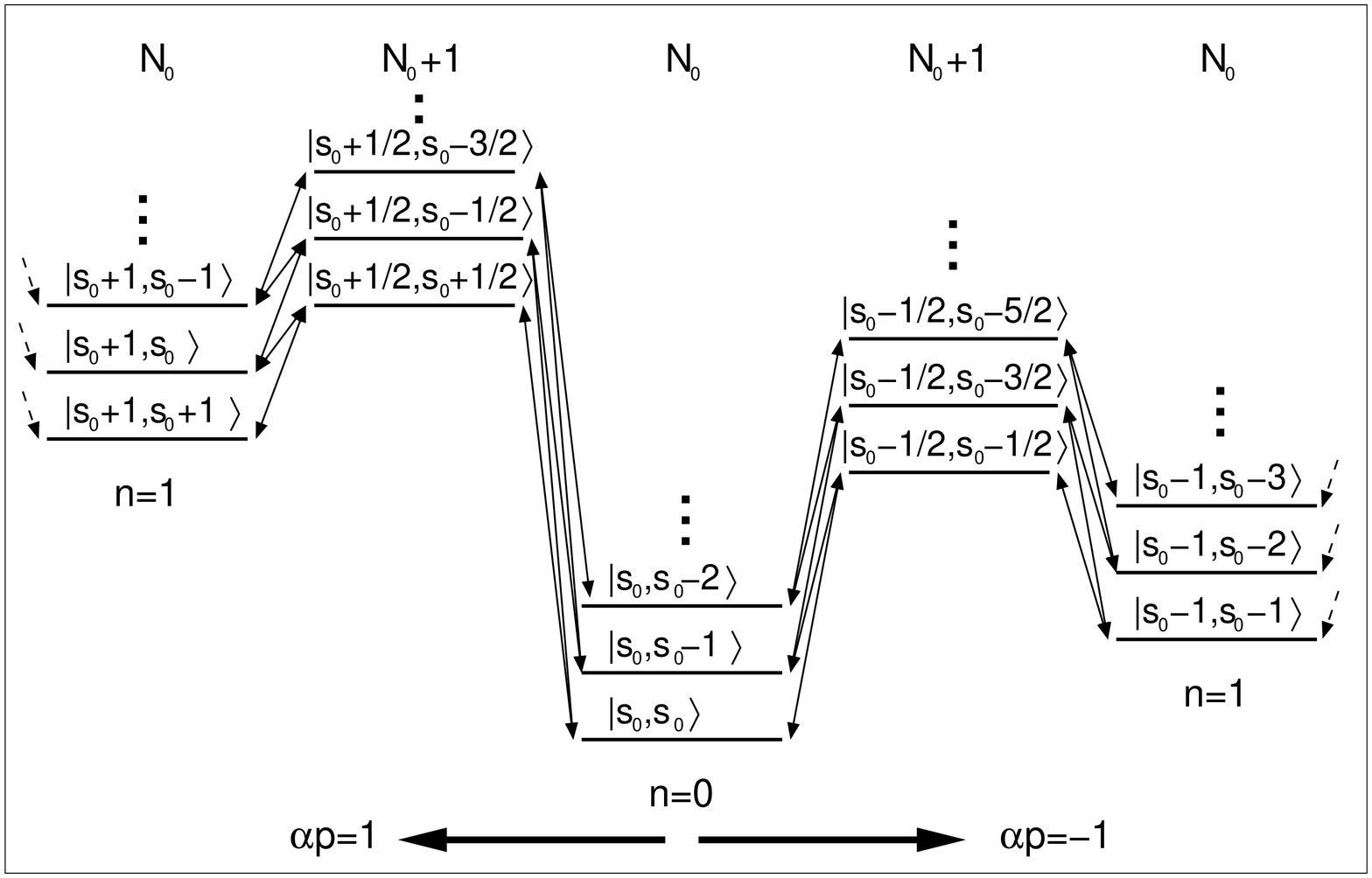}}
\caption{\label{fig:spin_wave_ladder} Illustration of a
nonequilibrium scenario involving spin-wave states: Different
multiplets $|s,m\rangle$ for $N_0$ and $N_0+1$ electrons on the grain
are depicted. Due to selection rules only specific transitions
(marked by arrows) are possible. }
\end{figure}

\subsection{\label{sec:nonequilibrium_spin_wave}Spin wave excitation}
In the preceding section,
we considered only transitions between different spin ground
states, $\ket{s,s}$; we shall now extend this scenario to
include {\em all} higher lying states of the corresponding
multiplets $\ket{s,m}$ (see Fig.~\ref{fig:spin_wave_ladder}). We
shall show that this results in a fine structure for the current
steps which would be resolvable only for {\em very} low
temperatures\cite{temperature}
 and hence would not be expected to be observable
 in DGR's present measurements. Moreover, we find that the Zeeman
splitting of resonances is strongly suppressed for a large spin
$s_0 \gg 1$.

Fig.~\ref{fig:spin_wave_ladder} illustrates the  transitions between
different spin wave states of ``neighboring'' multiplets.
For simplicity, we assume that both the
magnetic field  and  the easy axis
lie in the $z$-direction,
so that the selection rule $|m_f-m_i|=1/2$ holds.
\begin{figure}[t]
{\includegraphics[clip,width=0.98\linewidth]{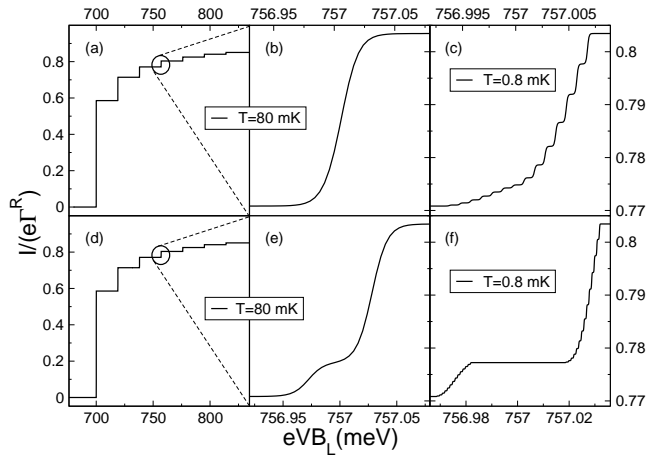}}
\caption{\label{fig:spin_waves}Current as function of $eVB_L$ for
the following parameters: $\alpha ={\textrm{min}}$, $p=+1$,
$s_0=10$, ${d}_{\textrm{min}}=1.19\, \textrm{eV}/s_0$,
${d}_{\textrm{maj}}= 4.61\, \textrm{eV}/s_0$, $U=2\,
\textrm{eV}/s_0$, $B_L=0.4$, $B_R=0.6$, $k_N=0.01$ meV,
$\Gamma^L/\Gamma^R=0.8$, $\Gamma^{\textrm{sf}}=0$,
$\Gamma^{\textrm{el}}=0$ and a Coulomb blockade region of  $7\,
\textrm{eV}/s_0$;
 (a-c): $h=0$ meV; (d-f):
$h=0.05$ meV; (b,e) shows the fourth current step of (a,d) for $T=80$mK, (c,f) shows
the same step for a lower temperature of $T=0.8$ mK.
No significance should be attached to step heights here,
since they depend on (unknown) tunneling matrix elements.}
\end{figure}

We  solved Eq.~(\ref{eq:master_equation}) numerically
for a spin\cite{smallspin} of $s_0=10$ and a Coulomb blockade
region of $7\,\textrm{eV}/s_0$ [as in Fig. 2, sample 3 of
Ref.~\onlinecite{gueron}], and calculated the current
 for zero magnetic field [Fig.~\ref{fig:spin_waves}(a-c)], and
non-zero applied field [Fig.~\ref{fig:spin_waves}(d-f)].

Fig.~\ref{fig:spin_waves}(a) shows the current as function of
energy ($B_LeV$) for a temperature of $T=80$ mK and zero magnetic
field, $h=0$ eV. The current displays seven equally spaced steps.
These steps belong to transitions between successive sets of
pairs of multiplets, \textit{e.g.}\  the first one belongs to
transitions between $\ket{s_0,m}$ and $\ket{s_0-1/2,m'}$, the next
one to transitions between $\ket{s_0-1,m}$ and
$\ket{s_0-3/2,m'}$, etc. Fig.~\ref{fig:spin_waves}(b) shows the
fourth current step of (a) on a finer energy scale. In
Fig.~\ref{fig:spin_waves}(c) the same step is shown for a lower
temperature of $T=0.8$ mK, at which it now reveals substructure
in the form of 14 finer steps. These {\em small} steps correspond
to  transitions between various states of the two multiplets,
$\ket{s_0-3,m}$ and $\ket{s_0-7/2,m}$, namely the first {\em
small} step in (c) corresponds to a transition, $\ket{s_0-3,\pm
(s_0-4)}\rightarrow\ket{s_0-7/2,\pm (s_0-7/2)}$, the next one to
$\ket{s_0-3,\pm (s_0-5)}\rightarrow\ket{s_0-7/2,\pm (s_0-9/2)}$,
etc. The last step in (c) corresponds to $\ket{s_0-3,\pm
(s_0-3)}\rightarrow\ket{s_0-7/2,\pm (s_0-7/2)}$. It can be
checked  easily, using Eq.~(\ref{Huni}), that their spacing is
given by  $k_N/s_0$ $[\simeq 0.01\,\textrm{meV}/s_0]$. Similarly,
all other steps of (a) (except for the first one) have a
substructure of smaller steps belonging to all transitions
between neighboring multiplets that are allowed by the selection
rules (as indicated in Fig.~\ref{fig:spin_wave_ladder}). Note
that the number of (large-scale) steps in (a) is still given by
Eq.~(\ref{eq:n_max}) of
Sec.~\ref{sec:nonequilibrium_spin_ground_states}, with a spacing
given by  $ d_{\textrm{min}}-U/2$ $ [=0.2\,\textrm{eV}/s_0]$.

\begin{figure}[t]
{\includegraphics[clip,width=0.98\linewidth]{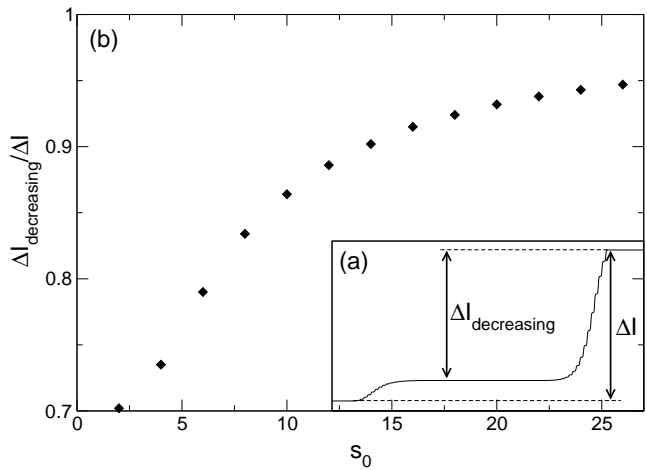}}
\caption{\label{fig:Zeeman} (a) A Zeeman-split current
step, where $\Delta I_{\textrm{decreasing}}$ is the overall
height of steps due to all $S^z$-decreasing transitions and
$\Delta I = \Delta I_{\textrm{decreasing}} + \Delta
I_{\textrm{increasing}}$ is the overall current step.
(b) $\Delta I_{\textrm{decreasing}}/\Delta I$, calculated
numerically for various $s_0$-values for the second step in the
current and $\Gamma ^{L}/\Gamma ^{R}=0.1$. }
\end{figure}

In Figs.~\ref{fig:spin_waves}(d) to (f) a similar set of plots is
shown as in Figs.~\ref{fig:spin_waves}(a) to (c), but now in the
presence  of an  applied magnetic field, $h=0.05$ meV.
 Fig.~\ref{fig:spin_waves}(d) shows the current itself.
Figs.~\ref{fig:spin_waves}(e) and (f) again show the
fourth current step for two different temperatures.
Fig.~\ref{fig:spin_waves}(e) shows that the step of (b)
has Zeeman-split into two steps and (f) shows that these
two steps correspond to two groups of transitions, namely
all $S^z$-{\em increasing} transitions
(steps on the left side) and $S^z$-{\em decreasing} transitions (right side)
between the two  multiplets $\ket{s_0-3,m}$ and $\ket{s_0-7/2,m}$.
The fact that (f) shows many more steps than (c) results
from the fact that the applied magnetic field
lifts the degeneracy of states $\ket{s,\pm m}$.
The first step in (f) belongs to a transition
$\ket{s_0-3,s_0-4}\rightarrow\ket{s_0-7/2,s_0-7/2}$ the next one to
$\ket{s_0-3,s_0-5}\rightarrow\ket{s_0-7/2,s_0-9/2}$, etc.
The last step corresponds to
$\ket{s_0-3,s_0-3}\rightarrow\ket{s_0-7/2,s_0-7/2}$.

\begin{figure}[t]
{\includegraphics[clip,width=0.57\linewidth]{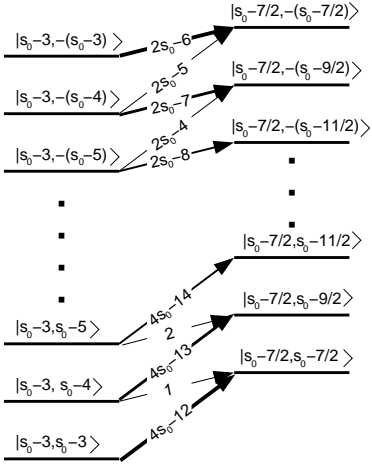}}
\caption{\label{fig:stepheight} 
Transitions from the  ($s_0-3$) to the ($s_0-7/2$)-multiplet.
Arrwos indicate all transitions allowed by the selection rule that $S^z$ can change only by
$\pm 1/2$. The width of an arrow schematically indicates the size of the 
Clebsch-Gordan coefficient involved in the matrix element of the corresponding transition.
The energy separations between the various levels are not drawn to scale. 
The order in which transitions become possible as the applied voltage is increased
is given by the numbers in the arrows, which
range from $1$ to $2s_0-6$ for $S^z$-increasing transitions, and then from 
$2s_0-5$ to $4s_0-12$ for $S^z$-decreasing transitions.}
\end{figure}

Since the individual substeps in
Fig.~\ref{fig:spin_waves}(f) are higher for the $S^z$-decreasing
transitions (to the right of the plot) than for the
$S^z$-increasing transitions (to the left), the second
large-scale step in Fig.~\ref{fig:spin_waves}(e)
 is higher than the first. The reason for this height
difference lies in the fact that matrix elements for transitions
between different spin wave states contain Clebsch-Gordan
coefficients: Let us consider the two multiplets which give rise to 
the current steps in Fig.~\ref{fig:spin_waves}(f), namely the ($s_0-3$) and the ($s_0-7/2$)-multiplet.
The energy levels of these multiplets are schematically depicted in Fig.~\ref{fig:stepheight}.
The first transition between the multiplets 
which becomes possible as the applied voltage is increased is a $S^z$-increasing transition,
namely the 
$\ket{s_0-3,s_0-4}\rightarrow \ket{s_0-7/2,s_0-7/2}$-transition at the bottom of the ladder.
As the voltage is increased further more and more  $S^z$-increasing transitions become possible.
(In Fig.~\ref{fig:stepheight} they are numbered as $1$,$2$,$\cdots$,$2s_0-6$.) 
The last $S^z$-increasing transition, namely $\ket{s_0-3,-(s_0-3)}\rightarrow\ket{s_0-7/2,-(s_0-7/2)}$,
lies at the top of the ladder. As we increase the applied voltage further the first 
$S^z$-decreasing transition ($\ket{s_0-3,-(s_0-4)}\rightarrow\ket{s_0-7/2,-(s_0-7/2)}$)
becomes possible (at the top of the ladder). As we increase the voltage further
more and more $S^z$-decreasing transitions become possible, the last transition 
($\ket{s_0-3,s_0-3}\rightarrow\ket{s_0-7/2,s_0-7/2}$)
lying at the bottom of the ladder. 
Let us now compare Clebsch-Gordan coefficients involved in the 
different transitions. In Fig.~\ref{fig:stepheight} their magnitude 
is schematically indicated by the width of the arrows marking the transitions.
For example, the Clebsch-Gordan coefficient involved
in the matrix element for the first $S^z$-increasing transition
($\ket{s_0-3,s_0-4}\rightarrow\ket{s_0-7/2,s_0-7/2}$) as well as the 
first $S^z$-decreasing transition 
($\ket{s_0-3,-(s_0-4)}\rightarrow\ket{s_0-7/2,-(s_0-7/2)}$) in
Fig.\ref{fig:spin_waves}(f) (and Fig.\ref{fig:stepheight}) is of order $O(1/s)$, hence these
transitions carry very little weight. By increasing the applied voltage, 
so that
more transitions become accessible, the Clebsch-Gordan
coefficients for $S^z$-increasing  transitions 
increase from
$O(1/s)$ to $O(1)$ as we go \emph{up}  the ladder and those for 
$S^z$-decreasing transitions likewise increase from $O(1/s)$ to $O(1)$ 
as we subsequently go \emph{down} the ladder.
Now, the occupation probabilities
$P$, in general, are larger\cite{Clebsch-Gordan} for lower lying
spin wave states than for higher-lying ones. 
Since transitions out of these more strongly-populated lower lying states 
of the spin-($s_0-3$)-multiplet have  a Clebsch-Gordan coefficient of order $O(1)$ 
if they are $S^z$-decreasing, and of order $O(1/s)$ if they are $S^z$-increasing,
we conclude that the total weight of all $S^z$-decreasing transitions is larger than that
of all $S^z$-increasing transitions. The net result is
that the $S^z$-decreasing step in  Fig.\ref{fig:spin_waves}(e) is
higher than the $S^z$-increasing step.

The relative heights of  the two  large-scale steps in
Fig.~\ref{fig:spin_waves}(e) are analyzed in
Fig.~\ref{fig:Zeeman}, or more precisely,
 the height of the current step due to all $S^z$-decreasing
 transitions, $\Delta I_{\textrm{decreasing}}$,
 relative to the total height of the current step
 $\Delta I$. Fig.~\ref{fig:Zeeman}
 confirms that \emph{$S^z$-decreasing transitions typically
carry significantly more
 weight than $S^z$-increasing transitions.} Moreover, with
 increasing spin $s_0$ the height of $\Delta
 I_{\textrm{decreasing}}$ increases strongly relative to $\Delta
 I$,
 so that for large $s_0$ no Zeeman
splitting is expected.

Let us now summarize the consequences of the
nonequilibrium scenario discussed above for the tunneling spectra
measured by DGR. (i) Firstly, the temperature in DGR's
experiments, namely $T \approx 80$~mK, is too large for the fine
current steps of Fig.~\ref{fig:spin_waves}(c) and (f) due to spin
wave excitations to have been observable. Instead, only the
large-scale current steps of Fig.~\ref{fig:spin_waves}(a) and (d)
would be observable. The observed resonance spacing of $\delta
E_{\textrm{res}}\approx 0.2$meV [cf. (P1)] indeed does agree with
that expected for minority-electron charging transitions and $s_0
\simeq 1000$ [cf. Eq.~(\ref{eq:spacing-estimates-min})].  (ii)
Secondly, the nonequilibrium scenario discussed above can also
account  for the fact (P3) that the vast majority of the observed
transitions within a given sample shift in energy with a similar
slope for large magnetic fields, since for $s_0 \gg 1$,
$S^z$-decreasing transitions carry far more weight than $S^z$-increasing
transitions [Fig.~\ref{fig:Zeeman}].

\subsection{\label{sec:nonequilibrium_single_particle}Spin
and single particle excitations}
\begin{figure}[t]
{\includegraphics[clip,width=0.98\linewidth]{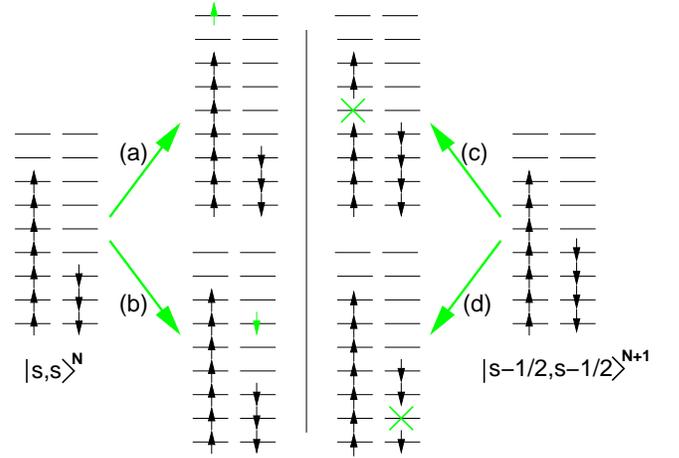}}
\caption{\label{fig:single_particle_illustration}
Illustration of single particle excitations:
Examples of excited single particle states, which can
be reached from the ground state $\ket{s,s}^{N}$ by (a)
a majority electron or (b) a minority electron
entering the grain, or from a $\ket{s-1/2,s-1/2}^{N+1}$-state
by (c) a majority electron or (d) a minority electron leaving the grain.}
\end{figure}
We saw in
Secs.~\ref{sec:nonequilibrium_spin_ground_states}
(and~\ref{sec:nonequilibrium_spin_wave}) that a nonequilibrium
scenario involving transitions between different spin states (spin
wave states) of the grain lead to resonances in tunneling spectra
spaced by $ d_{\textrm{min}}-U/2$, which for $s_0=1000$ gives
a value of $\approx 0.2$meV, as observed in experiments. However,
these scenarios are not always able to explain the large
number of resonances observed, since $n_{\textrm{res}}^{\textrm{ch}}$
given in Eq.~(\ref{eq:n_max}) depends strongly on the threshold
charging energy and can become as small as {\em two} when
$E_{\textrm{C}}^{\textrm{thresh}}$ is of order or smaller than
$ d_{\textrm{min}}$.

We  shall now argue below that the abundance of resonances
measured by DGR can be explained by taking the analysis one step
further, namely by including single-particle excitations in
addition to spin excitations. For the spin excitations we shall
henceforth restrict our considerations to transitions between
spin ground states as in
Sec~\ref{sec:nonequilibrium_spin_ground_states}, since we saw in
Sec.~\ref{sec:nonequilibrium_spin_wave} that DGR's experimental
temperature was too high to resolve the fine structure due to
spin wave transitions. Furthermore, we shall
assume\cite{rates}
$\Gamma^{\textrm{el}}\gg\Gamma^{\textrm{tun}}$, and hence shall
take into account only excited single particle states involving a
\emph{single} particle-hole pair, \textit{i.e.}\ states which can
be reached from the corresponding ground states by a {\em single}
tunneling transition, namely a majority/minority electron
entering the $N$-electron grain  or a majority/minority electron
leaving the $N+1$-electron grain.~\cite{single_particle}
Fig.~\ref{fig:single_particle_illustration} illustrates some
examples.

\begin{figure}[t]
{\includegraphics[clip,width=0.98\linewidth]{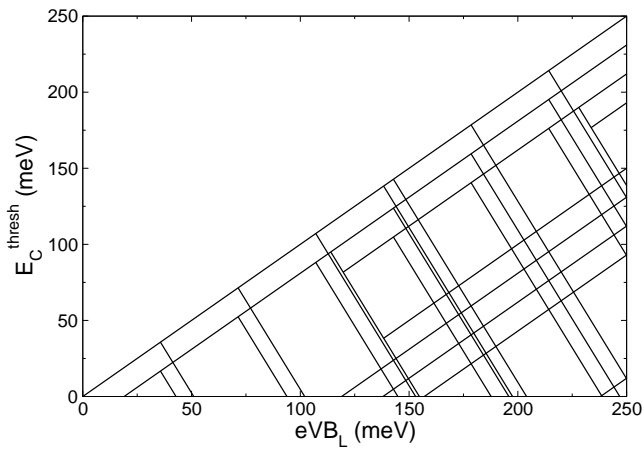}}
\caption{\label{fig:E_C^thresh}
 Resonances as function of
$E_{\textrm{C}}^{\textrm{thresh}}$ and $eVB_L$ for the same
parameters as Fig.~\ref{fig:single_particle}. As
$E_{\textrm{C}}^{\textrm{thresh}}$ is increased additional
resonances appear.}
\end{figure}

We solved Eq.~(\ref{eq:master_equation}) numerically for the
parameters of Fig. ~1 in Ref.~\onlinecite{mandar} and calculated
the current; the result is shown in
Fig.~\ref{fig:single_particle}(b). The current shows $16$ steps
in the plotted region. Note that the calculation of the current
was done for the same parameters as in
Fig.~\ref{fig:single_particle}(a) where only spin excitations
were taken into account. The arrows in (b) mark all additional
resonances with respect to (a) which arise due to a combination of
spin- and single particle excitations. Most of the
resonances in Fig.~\ref{fig:single_particle}(b) are due to
tunneling of minority electrons, namely $14$ resonances out of
$16$. This results from the very different density of states of
minority and majority electrons at the Fermi energy [cf. (P4)].

Thus, when  single-particle excitations are considered in
addition to nonequilibrium spin accumulation, additional
resonances appear at higher voltages, so that the number of
resonances increases significantly. The resonances are no longer
equally spaced, as was the case for pure spin
excitations, but the average spacing is of the same order of
magnitude.

We also investigated the gate-voltage dependence of these
resonances. Fig.~\ref{fig:E_C^thresh} shows resonances as
function of bias voltage
 ($eVB_L$) and threshold charging energy
$E_{\textrm{C}}^{\textrm{thresh}}$.
Note that
$E_{\textrm{C}}^{\textrm{thresh}}$ depends linearly on gate
voltage via Eq.~\ref{eq:defineECthresh}.
Fig.~\ref{fig:E_C^thresh} shows clearly that the number of
resonances increases with increasing
$E_{\textrm{C}}^{\textrm{thresh}}$ [cf.(P5)]. This is in
qualitative agreement with experimental data [Fig. 3(a)
in Ref.~\onlinecite{mandar}], although a detailed understanding of
the observed gate voltage dependence of resonances would
require a more systematic examination.

\section{\label{sec:conclusions}Conclusions}
In summary, we have shown that nonequilibrium
spin and single-particle excitations within
our model are able to explain  most of the
experimental data by DGR on tunneling spectroscopy of
ultrasmall ferromagnetic grains. Nonequilibrium
spin excitations lead to a resonance spacings
 and an absence of Zeeman splitting as observed in experiments.
By taking single-particle-excitations into account in addition to
spin excitations, it is possible to satisfactorily
explain the large number of resonances observed in
experiment.

\begin{acknowledgments}
We thank C. Canali and A. MacDonald for advance communication of
their work and several very helpful discussions, and J.  Becker,
D. Boese, A. Brataas, E.  Chudnovsky, A. Garg, C. Henley,  D.
Loss, W. M\"orke, A. Pasupathy, J. Petta, and G. Sch\"on for
fruitful discussions. Special thanks to D.C. Ralph, M. Deshmukh,
E. Bonet and S. Gu\'eron for a fruitfull collaboration in which
they not only made their data available to us but also
significantly contributed to the development of the theory.  This
work was supported by the DFG through SFB195, the DFG-Program
``Semiconductor and Metallic Clusters'' and by the DAAD-NSF.
\end{acknowledgments}


\begin{thebibliography}{}
\bibitem{bulkN} We use ${d}_{\sigma}=1/[N_{\textrm{a}} {\cal N}_{\sigma}(\varepsilon_{\textrm{F}})]$,
where ${\cal N}_{\sigma}(\varepsilon)$ is the spin-dependent bulk density of states
per atom.~\cite{Papa} Using bulk values is appropriate for nanoscale grains,
    since quantum chemical
    calculations~\cite{rosch} on Ni clusters have shown that ${\cal N}_\sigma
    (\varepsilon)$ has converged to its bulk shape already for $N_{\textrm{a}}$ as
    small as 6. For Co, Ref.~\onlinecite{Papa} gives
    ${\cal N}_\uparrow (\varepsilon_{\textrm{F}})=0.18$eV$^{-1}$, ${\cal
N}_\downarrow (\varepsilon_{\textrm{F}})=0.7$eV$^{-1}$.
\bibitem{CMcD} A model similar in spirit was
introduced independently by Canali and MacDonald.~\cite{canali}
\bibitem{anisotropy}
A more general form for the anisotropy of an
ultrasmall ferromagnetic grain would be
$ {\cal H}_{\textrm{anis}} =
  - \: \sum_{ab} \sum_{ij} S^a_i {\cal K}^{ab}_{ij} S^b_j$
where ${\cal K}^{aa}_{ij}$ is a hermitian, traceless tensor
[$\sum_a {\cal K}^{aa}_{ij} = 0$], which describes the energy cost for
rotating the various spins $\vec S_j$ [see Ref.~\onlinecite{prl_kleff}].
\bibitem{canali-mcd}
This is in agreement with Fig. 2 of Ref.~\onlinecite{canali}.
\bibitem{rates}
For DGR's Co nanograins of Ref.~\onlinecite{gueron} or
Ref.~\onlinecite{mandar}, we estimate $\Gamma^{\textrm{tun}}$ from
the conductance to be $\Gamma^{\textrm{tun}}\approx G
  d_{\textrm{min}} / e^2 \approx 10^9 {\textrm{s}}^{-1}$
  or $10^7 {\textrm{s}}^{-1}$, respectively.
  By comparison, we estimate
  the rate for spin-conserving electronic relaxation
 to be $\Gamma^{\textrm{el}}\approx 10^9 {\textrm{s}}^{-1}$,
 if we use Eq.(8)  of Ref.~\onlinecite{agam} (though it
 is not entirely clear that that formula, derived
 for ordinary metals, is also applicable to a strongly correlated
ferromagnetic system such as Co). Thus, for the grains of
Ref.~\onlinecite{gueron} or Ref.~\onlinecite{mandar},
$\Gamma^{\textrm{el}} $ is of the same order or larger than
$\Gamma^{\textrm{tun}}$. As to the spin-flip rate
$\Gamma^{\textrm{sf}} $, we do not know of a reliable way to
estimate it from theoretical considerations, except to state that
it is expected to be much smaller than the non-spin-flip relaxation rate,
$\Gamma^{\textrm{sf}} \ll \Gamma^{\textrm{el}}$. The condition
for the occurrence of nonequilibrium spin accumulation, namely
$\Gamma^{\textrm{tun}} > \Gamma^{\textrm{sf}}$, would thus
readily be satisfied for Ref.~\onlinecite{gueron}, but for
Ref.~\onlinecite{mandar}, one would need $\Gamma^{\textrm{sf}}$ to
be more than two  orders of magnitude smaller than our estimate for
$\Gamma^{\textrm{el}} $. This difference is rather large,
but might also indicate that our estimate for
$\Gamma^{\textrm{el}} $  is unreliably large.
\bibitem{no_hanis_hzee}
We do not consider ${\cal H}_{\textrm{uni}}$ or ${\cal
H}_{\textrm{Zee}}$ here, since  in DGR's experiments the
relevant energy scales, $k_N\approx 0.01$meV and
$h_{\textrm{max}}\approx 0.5$meV, are much smaller than the
exchange splitting of majority and minority bands,
$\Delta_{\textrm{F}}\approx 2$eV in Co.
\bibitem{negativ_spacing}
For general values of $d_{\textrm min}$, $d_{\textrm maj}$
and $U$, one of the resonance spacings in $d_\alpha - U/2$ in
Eq.~(\ref{eq:spacing}) could become negative. However, only
positive resonance spacings lead to observable conductance
resonances. Thus, if $d_{\textrm min} - U/2 <0$ (or $d_{\textrm
maj} - U/2 <0$), only transitions involving the tunneling of
majority (or minority) electrons would lead to observable
conductance resonances. Note, though,  that the stability
condition Eq.~(\ref{eq:inequality}) precludes both energy
differences being negative at the same time.
\bibitem{smallspin}
The ground state spin of $s_0 = 10$ chosen here is of
course unrealistically small, but for larger values the numerical
analysis becomes intractably complicated. We do believe, though,
that the conclusions drawn from our numerical analysis are
generic.
\bibitem{single_particle}
For simplicity we have not included excited single-particle
states which can only be reached from ground states by multiple
(nonequilibrium) excitations. However, these transitions
certainly exist as
$\Gamma^{\textrm{el}}\approx\Gamma^{\textrm{tun}}$ (see
Ref.~\onlinecite{rates}) and would lead to additional resonances.
Our aim, however, is to show that even by including only part of
the single-particle-excitations possible, the number of
resonances is increasing significantly [with respect to
$n_{\textrm{res}}^{\textrm{ch}}$ given by Eq.~(\ref{eq:n_max})], making it
possible to explain the large number of resonances in DGR's
measurements.
\bibitem{temperature}
For a spin of $s_0=1000$ this fine structure would only be resolvable for
$T<0.2$mK. This follows from a comparison of the spacing of the fine structure steps,
$k_N/s_0$, with $k_B T$.
\bibitem{Clebsch-Gordan}
Though this statement is of course somewhat of an oversimplification,
since the occupation probabilities depend in a sensitive
way on the transition matrix elements, we do believe that
it is, ultimately, the reason why in our numerical
calculations [Fig.~\ref{fig:Zeeman}] the total weight
of $S^z$-decreasing transitions turns out to be larger than for
$S^z$-increasing transitions. Note also, that in the calculation for Fig.\ref{fig:spin_waves} no relaxation
processes were taken into account, which of course is not realistic. In particular,
in the presence of relaxation it is to be expected that states with negative 
$S^z$-eigenvalues, which, loosely speaking correspond to a spin-vector
pointing in a direction opposite to the applied magnetic field, would have little or no 
occupation probability.
However, we believe that 
our qualitative discussion of the role of Clebsch-Gordan coefficients above 
would still be  valid for non-zero relaxation since a non-zero relaxation
would simply lead to an even larger occupation probability for the low lying spin wave states
and smaller occupation probability for the high lying states. 
Therefore, we expect the inclusion of   non-zero relaxation to make the ratio
$\Delta I_{\textrm{decreasing}}/\Delta I$ even larger and the width
of the plateau between the $S^z$-increasing and $S^z$-decreasing steps in 
Fig.\ref{fig:spin_waves}(e) wider.
\end{thebibliography}
\end{document}